\documentclass{ws-p8-50x6-00}
\usepackage{natbib, epsfig, graphics}
\def\newblock{}
\begin{document}

\title{Limits on Neutrino Masses from Large-Scale Structure}

\author{Eric Gawiser}

\address{U.C. San Diego, CASS/0424, 9500 Gilman Dr., La Jolla CA 92093-0424, 
USA\\E-mail: egawiser@ucsd.edu}

\maketitle

\abstracts{Massive 
neutrinos have a detectable effect on cosmological structure 
formation, in particular on the large-scale distribution of galaxies.  
Adding Hot Dark Matter to the now-standard $\Lambda$CDM model 
leads to a worse fit to large-scale structure and CMB anisotropy data.  
This results in a limit on the mass of the most massive neutrino of 
4 eV, assuming a power-law primordial power spectrum.}

\section{What Can Cosmology Tell Us About Neutrino Masses?}  

\hspace{0.2in}
There is now evidence for neutrino oscillations 
from SuperKamiokande~\cite{fukudaetal98} 
($\Delta m^2 \simeq 10^{-3}~$eV$^2$), 
the solar neutrino problem~\cite{bahcallks98} 
($\Delta m^2 \simeq 10^{-5}~$eV$^2$)\footnote{for 
vacuum oscillations, $\Delta m^2 \simeq 10^{-10}~$eV$^2$}
and LSND~\cite{athanassopoulosetal96}
($0.1$ eV$^2 \leq \Delta m^2 \leq 20~$eV$^2$). 
While these oscillations only test the difference in squared masses, they 
give evidence that the mass of at least one (and probably at least three) 
neutrino species is non-zero.  
Laboratory limits from tritium beta decay 
rule out the possibility of an electron 
neutrino more massive than 4.4 eV~\cite{belesevetal95}.   
Present cosmological bounds on the masses of other neutrino species are 
stricter than those from laboratory experiments; a 45 eV neutrino 
would lead to $\Omega_\nu=1$~\cite{kolbt90}, so for a universe at less than 
critical density the neutrinos must all be lighter than this.  
The exception to this is if a neutrino is so massive ($>1$ MeV) 
that it was non-relativistic during freeze-out, i.e. 
Cold Dark Matter (CDM).

	Each model of structure formation predicts Cosmic Microwave Background 
(CMB) anisotropy and large-scale structure inhomogeneities.  Massive neutrinos 
lead to slightly different predictions for CMB 
anisotropies and significantly 
different predictions for large-scale structure.  
The CMB radiation power spectrum is given by $C_\ell = \int_k 
C_\ell (k) P_p(k)$ and the matter power spectrum by $P(k) = 
T^2(k)P_p(k)$, where $C_\ell (k)$ and $T(k)$ are transfer functions predicted by a given model of structure formation and $P_p(k)$ is the primordial power 
spectrum of density perturbations that originated in the early universe.  
Massive neutrinos alter these transfer functions, so for a given $P_p(k)$ we 
can test a model with massive neutrinos by comparing its predictions to the 
observed $C_\ell$ and $P(k)$.  	
The effect is to exponentially damp the matter transfer 
function, $T(k)$, on
scales smaller than the neutrino free-streaming scale\cite{huet98},  
\begin{equation}
k \simeq 0.026 \left(\frac{m_\nu}{1eV}\right)^{\frac{1}{2}} \Omega_m^{1/2} 
h ~ Mpc^{-1},
\end{equation}
which is equal to the horizon size when the neutrinos become non-relativistic.
  For
mostly Cold Dark Matter and a fraction of Hot Dark Matter 
(massive neutrinos), 
the damping is no longer exponential but still quite significant.  
The effect on the CMB is more subtle;
relativistic neutrinos increase the radiation density before decoupling, 
which affects the shape of the acoustic peaks in the CMB angular 
power spectrum.

\section{Approach}

\hspace{0.2in} 
We assume here that  
$\Lambda$CDM is the correct model of structure formation and 
that the primordial power spectrum is well-described by a power-law. 
We start with a version of $\Lambda$CDM which is in 
good agreement with 
observations of Type Ia supernovae, the cluster baryon fraction, the 
primordial deuterium abundance, 
and Hubble's constant,
with $\Omega_m=0.4$, 
$\Omega_b=0.04$, and $h=0.7$.\footnote{Smaller 
values of $\Omega_m$ will lead to 
tighter limits on $\Omega_\nu$.}   
However, this $\Lambda$CDM model is not a great fit to 
large-scale structure data~\cite{gawisers98}, so we investigate whether 
adding a Hot Dark Matter component will improve the fit.

	Our data compilation includes 
COBE and smaller-scale CMB anisotropy detections, 
measurements of $\sigma_8$ from galaxy 
clusters \cite{vianal96, bahcallfc97},
a measurement of the matter power spectrum from 
peculiar velocities
\cite{kolattd97},
redshift-space matter power spectra from 
SSRS2+CfA2
\cite{dacostaetal94},
LCRS 
\cite{linetal96},
PSCz 
\cite{sutherlandetal99}
and APM Clusters 
\cite{tadrosed98},
and a real-space matter power spectrum derived from the 
APM angular correlation function 
\cite{eisensteinz99}.
Using all available CMB and large-scale structure data 
gives us information on intermediate scales, which helps to differentiate 
between variations in the primordial power spectrum and the reduction in 
small-scale power caused by massive neutrinos.  We analyze this 
data compilation using the methods of
Gawiser \& Silk~\cite{gawisers98}.

Even given all of this data, we need to assume something about the 
primordial power spectrum.  
We have tried using Harrison-Zel'dovich (scale-invariant, $P_p(k)=Ak$) 
and scale-free ($P_p(k)=Ak^n$) primordial power spectra. 
For inflationary models, the most reasonable 
parameterization is 
$\log P_p(k) = \log A + n \log k + \alpha (\log k)^2 + ...$
with successive terms decreasing in importance in the slow-roll regime.  
Unfortunately, an unconstrained primordial power spectrum $P_p(k)$ can easily 
mimic the effect that massive neutrinos have on 
the matter transfer function $T(k)$, making it nearly impossible to 
limit the neutrino mass.

\section{Results}

\begin{figure}[htb]
\centerline{\psfig{file=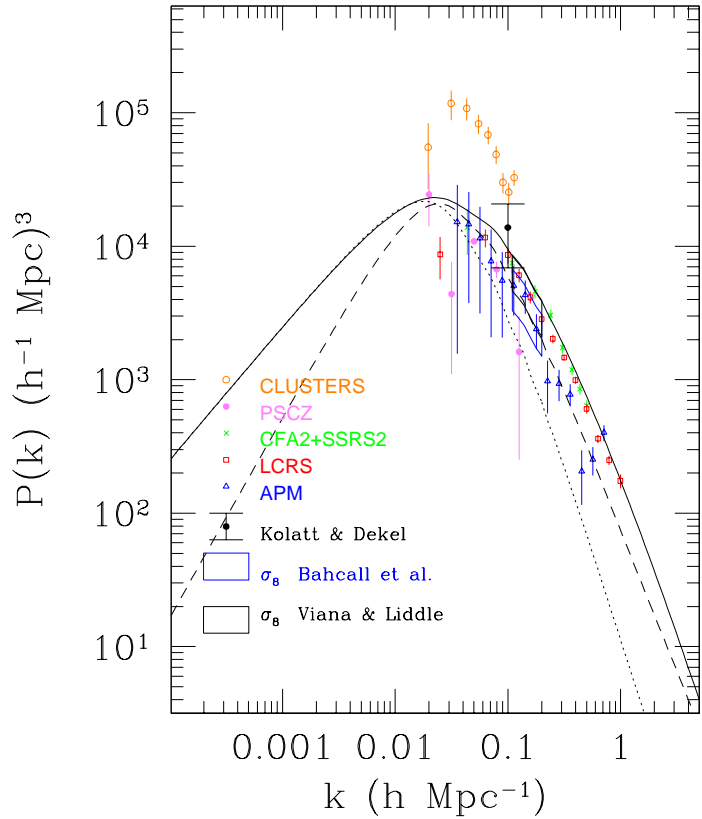,width=4.5in}}
\caption{Comparison of large-scale structure
 data compilation with $\Lambda$CDM (solid),  
$\Lambda$CHDM with $\Omega_\nu = 0.1$ and $n=1.0$ (dotted), and 
$\Lambda$CHDM with $\Omega_\nu = 0.1$ and $n=1.5$ (dashed).  Corrections 
for redshift distortions and non-linear evolution are not shown 
but were used to get quantitative results on scales $k \leq 0.2$.  
}
\label{fig:lchdm}
\end{figure}

\hspace{0.2in}
The results presented here were first determined by Gawiser\cite{gawiser99}.  
We find that as 
HDM is added, the combined fit to CMB and large-scale structure deteriorates.
This occurs because adding HDM reduces the power on physical 
scales shorter than the neutrino free-streaming length, which 
degrades the fit to large-scale structure data
 (see Figure \ref{fig:lchdm}).
For $\Omega_\nu = 0.05$, a blue tilt of the 
primordial power spectrum ($n=1.3$) 
is necessary to counteract the damping 
of small-scale perturbations by free-streaming of the 
massive neutrinos.
Even with this best-fit value of $n$, the 
fit to the data is worse than with no HDM, because 
CMB observations disfavor such a high value of $n$.  
For a higher HDM fraction, an even higher value of $n$ is preferred 
($n=1.5$ for the $\Omega_\nu=0.1$ model of Figure \ref{fig:lchdm}), 
leading to an even worse fit to the CMB data.

Our limits on the neutrino mass are based upon an attempt to search 
the reasonable parameter space around this fiducial model to 
produce the best fit possible to the data for a given neutrino 
mass.  Since disagreement with CMB data is the main problem once 
a blue tilt is considered, we have tried to alleviate this by 
adding a significant tensor component or early reionization.  Each of these 
effects reduces the small-scale CMB power relative to COBE scales, which 
helps to reconcile $n>1$ with the CMB data.  However, no parameter 
combination helps enough to make $\Lambda$CHDM a better fit than 
the fiducial $\Lambda$CDM model, and this allows us to set 
an upper limit on the sum of the neutrino masses,  
$\Sigma m_\nu = 94 \Omega_\nu h^2$ eV.  
An upper limit on $\Omega_\nu$ leads to an upper limit on the mass of 
the most massive 
neutrino.  If the mass is split between multiple nearly-equal-mass neutrinos, 
the limit on the sum of the masses is tighter because, for example, 
two 1 eV 
neutrinos depress the power spectrum more than one 2 eV 
neutrino because they both become non-relativistic later.  

If $\Lambda$CDM is right, and $H_0$ is about 65 $h^{-1}$Mpc and $n=1$, then 
$\Omega_\nu \leq 0.05, m_\nu \leq 2$~eV.

If $n$ can vary, 
$\Omega_\nu \leq 0.1, m_\nu \leq 4$~eV.

If $P_p(k)$ is not a power-law (non-inflationary or a complicated two-field inflationary model), then all bets are off.  

This is 
compatible with the recent claim by 
Croft et al.~\cite{crofthd99} 
that the Lyman $\alpha$ 
forest power spectrum plus COBE 
limits the neutrino mass to 5 eV or less.  Our method 
appears more robust as the Lyman $\alpha$ 
forest power spectrum has an uncertain normalization and 
covers a narrower  
range of scales than our large-scale structure compilation; moreover,
the origin of the Lyman $\alpha$ forest is not yet well understood.
Fukugita et al. \cite{fukugitals00}
use only COBE and $\sigma_8$ and assume $n=1$, yielding $\sum m_\nu \leq 3$~eV.
This is highly model-dependent because the primordial power spectrum is 
nearly degenerate with neutrino free-streaming when structure formation 
is only probed at two different spatial scales.
$\Lambda$CHDM has also been explored by 
Valdarnini et al.~\cite{valdarninikn98} and Primack \& 
Gross~\cite{primackg98} 
with different analysis 
methods and significantly smaller data compilations. 

Hu et al.~\cite{huet98} discuss the future prospects of 
including the Sloan Digital Sky Survey (SDSS) $P(k)$ in a method similar 
to that used here; 
they expect to detect or limit $m_\nu$ down to 0.5 eV.  
Cooray~\cite{cooray99} gives an expected future limit from measuring 
$P(k)$ with surveys of weak gravitational lensing of 
$m_\nu \leq 3$eV.

\section{Conclusions}

\hspace{0.2in}
The currently-favored $\Lambda$CDM model does not prefer the addition 
of a Hot Dark Matter component.  This leads to an upper limit on the 
mass of the most massive neutrino of 4eV if a power-law primordial 
power spectrum is assumed.
This limit is comparable to tritium beta-decay limits on the electron neutrino
 mass, and it should improve considerably with data from SDSS and the 
MAP satellite.  
Our results are already in conflict with the portion of the parameter space 
compatible with 
LSND~\cite{athanassopoulosetal96} 
that requires a mass difference greater than $16$~eV$^2$.  

In evaluating these results and other work on this subject, the 
reader is encouraged to 
check what assumptions have been made about cosmological 
models, the selection and 
normalization of data, and the primordial power spectrum.  Restrictive 
assumptions can lead to tight but meaningless limits on the neutrino mass.


\section*{Acknowledgments}

\hspace{0.2in}
I would like to thank Joe Silk for initially suggesting an investigation of adding Hot Dark Matter to $\Lambda$CDM and for his continuing collaboration in 
this research.  








\end{document}